\documentclass{article}
\usepackage{spconf,amsmath,graphicx}
\usepackage{float}
\usepackage{subfigure}
\usepackage{algorithm,algorithmicx}
\usepackage{multicol}
\usepackage{multirow}
\usepackage{array}
\usepackage{cite}
\usepackage{threeparttable}
\usepackage{booktabs}
\usepackage{graphicx}
\usepackage{epstopdf}
\usepackage{epsfig}
\usepackage{xcolor}
\usepackage{amsmath}
\usepackage{amsthm}
\usepackage{amssymb}
\usepackage{color}
\usepackage{url}
\usepackage{bigstrut}
\usepackage[T1]{fontenc}
\usepackage{autobreak}
\usepackage{amsfonts}
\usepackage{bm}
\usepackage{algpseudocode}

\title{An invariant feature extraction for multi-modal images matching.}

\name{Chenzhong Gao\textsuperscript{1,2},  Wei Li\textsuperscript{1,2}\thanks{This paper is supported by the Beijing Natural Science Foundation (Grant no. JQ20021), National Natural Science Foundation of China (Grant nos. 61922013, 61421001 and U1833203). (Email: liwei089@ieee.org)}}
\address{\textsuperscript{1}School of Information and Electronics, Beijing Institute of Technology, Beijing, China\\
\textsuperscript{2}Beijing Key Lab of Fractional Signals and Systems, Beijing, China}

\begin{document}
\maketitle

\begin{abstract}
This paper aims at providing an effective multi-modal images invariant feature extraction and matching algorithm for the application of multi-source data analysis. Focusing on the differences and correlation of multi-modal images, a feature-based matching algorithm is implemented. The key technologies include phase congruency (PC) and Shi-Tomasi feature point for keypoints detection, LogGabor filter and a weighted partial main orientation map (WPMOM) for feature extraction, and a multi-scale process to deal with scale differences and optimize matching results. The experimental results on practical data from multiple sources prove that the algorithm has effective performances on multi-modal images, which achieves accurate spatial alignment, showing practical application value and good generalization.
\end{abstract}


\section{Introduction}
\label{sec:intro}

Image matching has been a classical, important, and challenging work in image processing, especially for multi-modal images. Nowadays, the focus of image matching or registration has shifted to multi-modal and multi-source techniques, which are more critical to the practical applications. Multi-modal image matching is an essential step to implement data fusion, change detection, collaborative classification, joint analysis, and other image technologies.

Image matching algorithms are generally divided into area-based methods and feature-based methods according to technical means \cite{zitova2003image}, in which the feature-based ones are the most widely applied nowadays. By locating keypoints in the same positions, and accurately matching them as far as possible through similar or invariant features, the spatial transformation between images is represented in a sparse way. This type of algorithm is rather a good solution to the automatic multi-modal images matching, which has been a hot spot in recent years.

Chen et al. \cite{chen2010partial} proposed partial intensity invariant feature descriptor (PIIFD) based on scale-invariant feature transform (SIFT) \cite{lowe1999object,lowe2001local,lowe2004distinctive} for multi-source retinal images, which reduces intensity distortion problems. PSO-SIFT \cite{ma2016remote} uses second order gradient and an enhanced matching method for multi-source remote sensing images. Ye et al. developed histogram of orientated phase congruency (HOPC) \cite{ye2017robust,ye2018local}, in which an extended phase congruency model is designed for feature description. Radiation-variation insensitive feature transform (RIFT) \cite{li2019rift} utilizes a maximum index map (MIM) as the feature map using LogGabor filter, which has high invariance to intensity distortion. Multi-scale histogram of local main orientation (MS-HLMO) \cite{gao2022multi} comprehensively considers the differences of multi-source images, and proposed a framework focusing on orientation feature, which achieves intensity, rotation, and scale robustness on multi-modal remote sensing images. The key problems have emerged from these studies: 1) intensity distortion, 2) rotation, 3) scale difference, which are the difficulties of multi-modal image matching and need to be solved. The feature extracted from images should be robust to the above three aspects as possible.

This research focuses on analyzing the invariance feature of multi-modal images, and a feature-based image matching algorithm is proposed, which uses phase congruency (PC) and Shi-Tomasi feature point for keypoints detection, LogGabor filter and a weighted partial main orientation map (WPMOM) for feature description, and a multi-scale process to deal with scale differences and optimize matching results. The experiments on remote sensing and medical data from multiple sources indicate that the proposed method has excellent, stable performance in multi-modal image registration, achieving accurate spatial alignment, which shows practical application value and good generalization ability.

\section{PROPOSED MATCHING METHOD}
\label{sec:method}

\begin{figure*}[!h]
 \centering
 \centerline{\includegraphics[width=13cm]{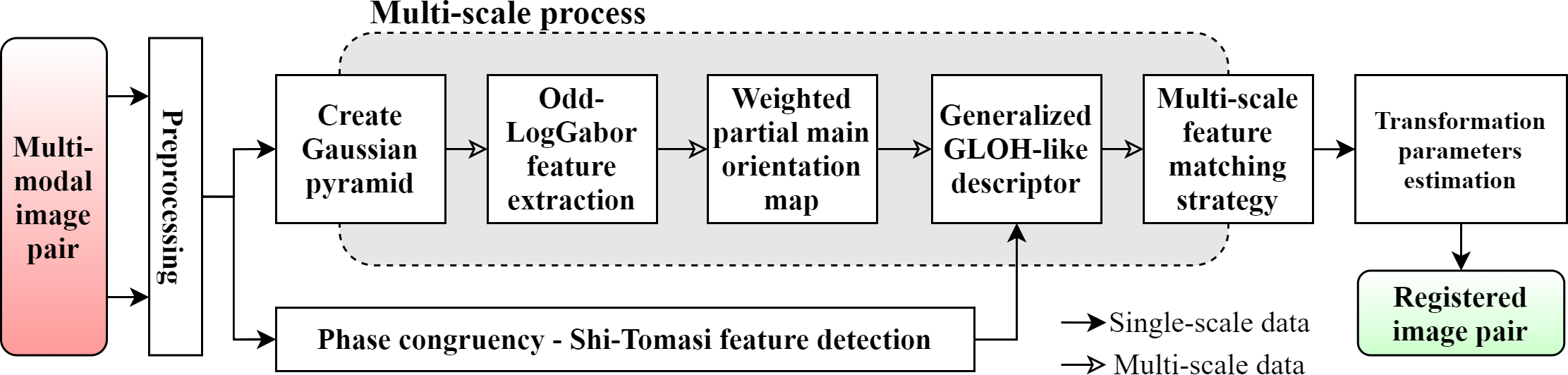}}
 \caption{The proposed multi-modal image matching framework.}
 \label{fig:framework}
\end{figure*}

The process of the proposed algorithm is shown in Fig.\ref{fig:framework}. The input multi-modal image pair are preprocessed. Then, keypoint detection based on phase congruency and Shi-Tomasi detector is performed. The feature extraction is carried out in a multi-scale process, in which Gaussian pyramids are first built. The odd-LogGabor filter is used to extract features of the images, then a weighted partial main orientation map (WPMOM) is calculated based on the LogGabor feature. The generalized GLOH-like descriptor (GGLOH) \cite{gao2022multi} is adopted to extract features for each keypoint. The keypoints are then matched through a multi-scale matching strategy. Finally, the spatial transformation is determined by the matched feature points.

\subsection{Feature points detection}
\label{ssec:subhead}
\begin{figure}[!h]
    \centering
        \subfigure[Optical-map pair]{\includegraphics[height=1.5in]{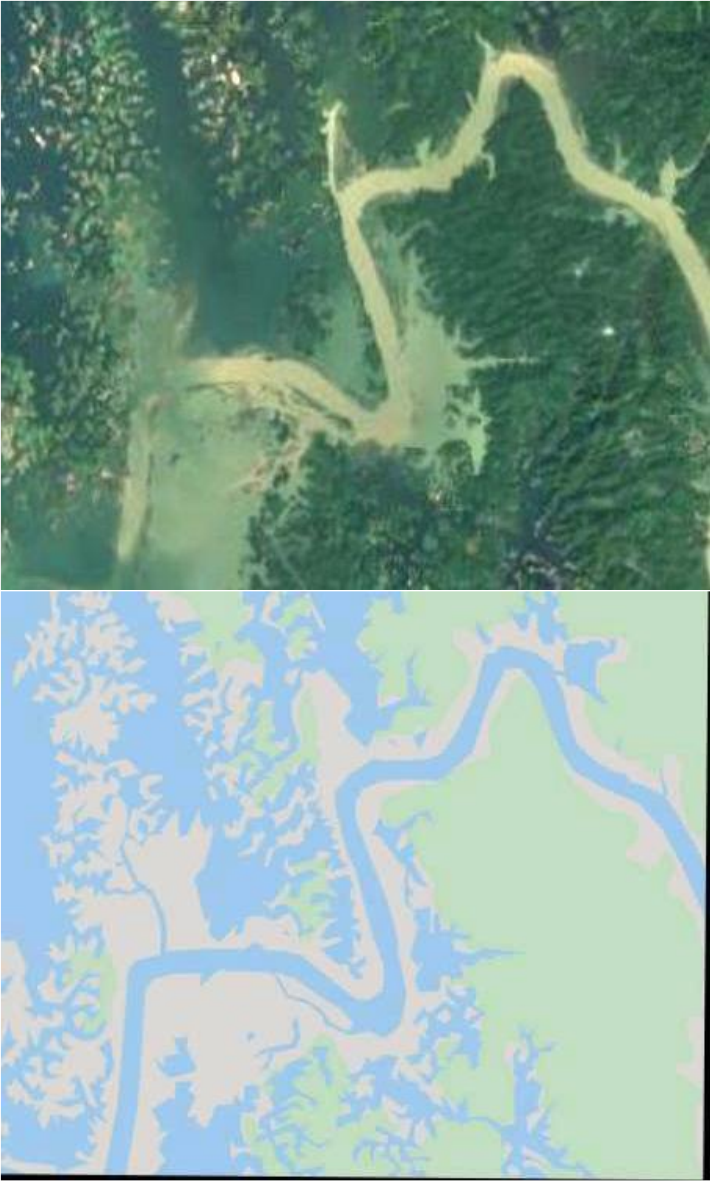}
        \label{fig:harris:a}}
        \subfigure[Harris points]{\includegraphics[height=1.5in]{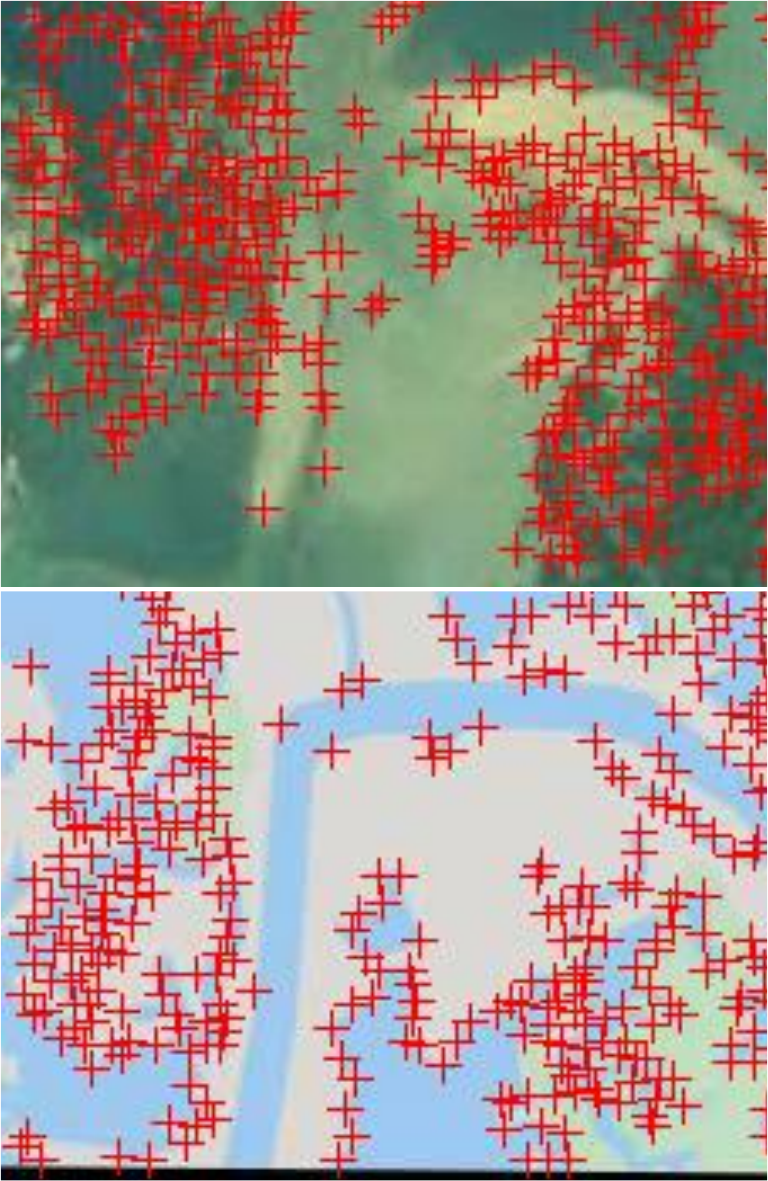}
        \label{fig:harris:b}}
        \subfigure[PC-Harris points]{\includegraphics[height=1.5in]{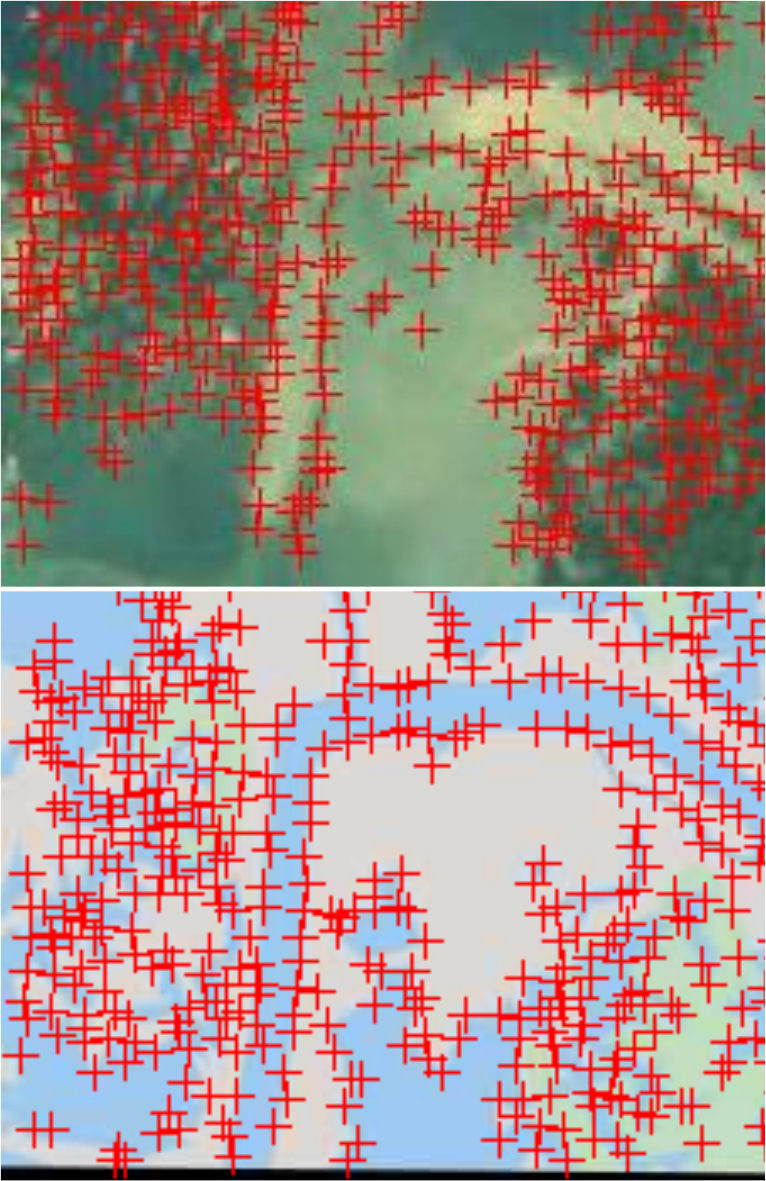}
        \label{fig:harris:c}}
    \caption{An example of multi-modal images feature detection.}
    \label{fig:harris}
\end{figure}

Harris corner detection \cite{harris1988combined} is an effective keypoints extraction method with stability and robustness, which has been widely used in multi-modal image matching. However, for multi-source images, the textures are likely to be very different, such as optical image and digital map as shown in Fig.\ref{fig:harris:a}. There are texture features in the optical image that do not exist in the digital map. Due to the influence of inconsistent texture information, the keypoints are difficult to locate in the same position, causing the repetition rate of points very low, as shown in Fig.\ref{fig:harris:b}. It is hoped that the extracted feature points are more focused on the salient and stable structure of the image. Thus the idea of adopting image's phase congruency is come up with.


Phase congruency (PC) has been proved to be robust under different imaging modalities \cite{ye2017robust,ye2018local,li2019rift}, which extracts stable structure features. The 2D-PC model can be calculated using components at multiple scales $s$ and orientations $o$ of the LogGabor wavelet \cite{kovesi2000phase,kovesi2003phase}:
\begin{equation} \label{eqKeyA}
{\bf{PC}}(x,y) = \frac{{\sum\nolimits_s {\sum\nolimits_o {{\omega _o}(x,y)\lfloor{{\bf{A}}_{s,o}}(x,y)\Delta {{\bf{\phi }}_{s,o}}(x,y) - T}\rfloor } }}{{\sum\limits_s {\sum\limits_o {{{\bf{A}}_{s,o}}(x,y)} }  + \varepsilon }}
\end{equation}
where ${\omega _o}(x,y)$ is the weighting factor based on frequency spread;  ${\bf{A}}_{s,o}$ is amplitude component of LogGabor response; $\Delta {{\bf{\phi }}_{s,o}}$ is the phase deviation; T is the noise compensation; $\lfloor\cdot\rfloor$ is a truncation function that produces the equal when positive and zero otherwise. A maximum moment map ${{\bf{M}}_\psi }$ which indicates edge features and a minimum moment map ${{\bf{m}}_\psi }$ which indicates corner features are then calculated as:
\begin{equation}
{\bf{a}} = \sum\nolimits_o {{{({\bf{P}}{{\bf{C}}_{{\theta _0}}}\cos ({\theta _0}))}^2}}
\end{equation}
\begin{equation}
{\bf{b}} = 2\sum\nolimits_o {({\bf{P}}{{\bf{C}}_{{\theta _0}}}\cos ({\theta _0})) \cdot ({\bf{P}}{{\bf{C}}_{{\theta _0}}}\sin ({\theta _0}))}
\end{equation}
\begin{equation}
{\bf{c}} = {\sum\nolimits_o {({\bf{P}}{{\bf{C}}_{{\theta _0}}}\sin ({\theta _0}))} ^2}
\end{equation}
\begin{equation}
{\bf{\psi }} = \frac{1}{2}\arctan (\frac{{\bf{b}}}{{{\bf{a - c}}}})
\end{equation}
\begin{equation}
{{\bf{M}}_\psi } = \frac{1}{2}({\bf{c}} + {\bf{a}} + \sqrt {{{\bf{b}}^2} + {{({\bf{a}} - {\bf{c}})}^2}} )
\end{equation}
\begin{equation}
{{\bf{m}}_\psi } = \frac{1}{2}({\bf{c}} + {\bf{a}} - \sqrt {{{\bf{b}}^2} + {{({\bf{a}} - {\bf{c}})}^2}} )
\end{equation}
where ${\bf{P}}{{\bf{C}}_{{\theta _0}}}$ is the PC map at orientation $\theta _0$. Then the two feature maps are added together to get a single map containing both edge and corner features:
\begin{equation}
{{\bf{M}}_w} = {{\bf{M}}_\psi } + {{\bf{m}}_\psi }
\end{equation}

Shi-Tomasi \cite{323794} feature is an improved algorithm of Harris, which states that the stability of the corner is related to the smaller eigenvalue of the feature matrix in Harris. Therefore, the Shi-Tomasi detector is performed on the feature map to extract stable feature points:
\begin{equation}
cornerness = min(\lambda_1,\lambda_2)
\end{equation}
\begin{equation} \label{eqKeyB}
\textbf{\emph{M}} = \left[ {\begin{array}{*{20}{l}}
{\sum\limits_{\bf{W}} {{{{\bf{M}}_w}_x}^2} }&{\sum\limits_{\bf{W}} {{{{\bf{M}}_w}_x}{{{\bf{M}}_w}_y}} }\\
{\sum\limits_{\bf{W}} {{{{\bf{M}}_w}_x}{{{\bf{M}}_w}_y}} }&{\sum\limits_{\bf{W}} {{{{\bf{M}}_w}_y}^2} }
\end{array}} \right]
\end{equation}
where $\lambda_1,\lambda_2$ are the eigenvalues of $\bf{M}$, ${{{\bf{M}}_w}_x}$ and ${{{\bf{M}}_w}_y}$ are the gradient of the weighted
moment map ${{\bf{M}}_w}$ in the $x$ and $y$ directions, and ${\bf{W}}$ is a gaussian window. After filtering the images using Eq.(\ref{eqKeyA})-(\ref{eqKeyB}), with local non-maximum suppression and threshold judgment, the keypoints are obtained from the multi-modal images, as shown in Fig.\ref{fig:harris:c}, the repetition rate of which is highly improved.

\subsection{Multi-modal robust feature extraction}
\label{ssec:subhead}
So far, there have been many studies on multi-modal image matching techniques, and a variety of robust features have been proposed, all of which have their effects and advantages. The key problem lies in finding the invariant or similar features among images with different modals. We reinspect the characteristics and matching methods of multi-modal image data and condense a core idea that is analyzed below to guide our feature design.

\begin{figure}[h!]
 \begin{center}
  \subfigure[]{\includegraphics[height=0.7in]{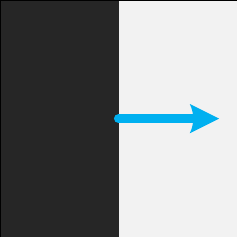}
  \label{fig:feat:a}}
  \subfigure[]{\includegraphics[height=0.7in]{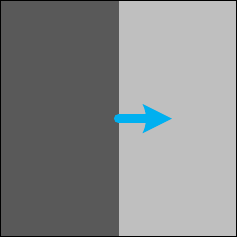}
  \label{fig:feat:b}}
  \subfigure[]{\includegraphics[height=0.7in]{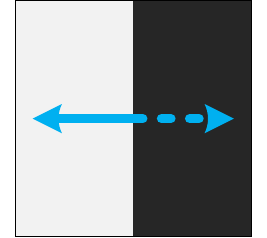}
  \label{fig:feat:c}}
  \subfigure[]{\includegraphics[height=0.7in]{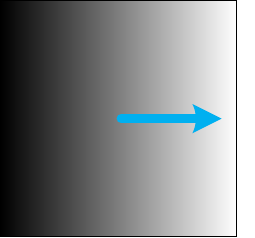}
  \label{fig:feat:d}}
  \caption{Local intensity distortions in multi-modal images and their robust gradient orientation features.}
  \label{fig:feat}
 \end{center}
\end{figure}

To intuitively show the characteristics of the multi-modal robust feature, we briefly summarize the common intensity distortion in multi-modal images, as shown in Fig.\ref{fig:feat}. The four images can be taken as the edge of an object or the interface of two substances. Assume that the center point is the keypoint, and the image block represents the intensity information of its neighborhood. In Fig.\ref{fig:feat:a}, the intensity amplitude on the left is lower than on the right. In Fig.\ref{fig:feat:b}, the gradient amplitude of the two parts changes. In Fig.\ref{fig:feat:c}, due to large intensity distortion, the gradient orientation is reversed. And Fig.\ref{fig:feat:d} represents a general situation, which is considered as non-linear intensity distortion or some degradation such as down-sampling or blurring. Notice that the gradient orientation remains on the same line, and for the reversion case, if the angle is limited to [-90°, 90°), the orientation is still 0°.

\begin{figure}[h!]
 \begin{center}
  \includegraphics[width=3.0in]{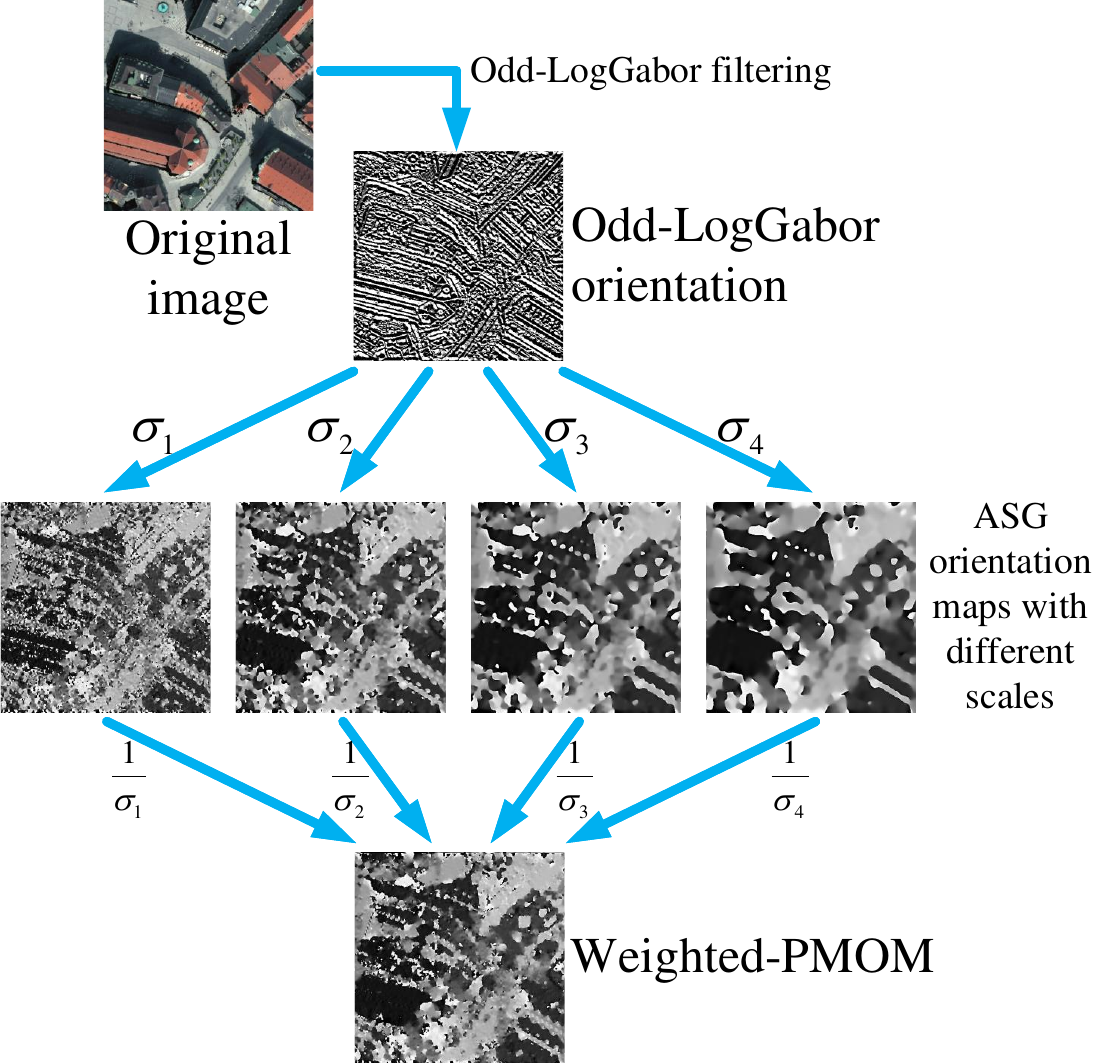}
  \caption{The establishing process of LogGabor-based Weighted-PMOM.}
  \label{fig:wpmom}
 \end{center}
\end{figure}

In multi-source images, due to differences in sensors, temporal, environments, etc., various intensity distortions may be caused, but the morphology of the most detailed part of the image basically lies in the four cases in Fig.\ref{fig:feat}. In summary, the magnitude of the local gradient varies, but the orientation is basically stable, which defines the feature information that should be focused on. In MS-HLMO \cite{gao2022multi}, a similar idea emerges, which prompts it to count only the gradient orientation in its PMOM feature map and abandon the amplitude. However, PMOM is still calculated using classic gradient that is very sensitive to image noise and inconsistency. So, in this research, the orientation is obtained with an odd-LogGabor filter. The odd-LogGabor wavelet has been widely used as a representation of gradient operator in image processing, which has much stronger robustness. The image's gradients along x and y directions based on odd-LogGabor are then calculated as:
\begin{equation}
\left\{ {\begin{array}{*{20}{c}}
{{\bf{G}}_x^{{\rm{LG}}}(x,y) = \sum\nolimits_s {\sum\nolimits_o {{\bf{I}}(x,y) * {\bf{LG}}_{s,o}^{{\rm{odd}}}(x,y) \cdot \cos (o)} } }\\
{{\bf{G}}_y^{{\rm{LG}}}(x,y) = \sum\nolimits_s {\sum\nolimits_o {{\bf{I}}(x,y) * {\bf{LG}}_{s,o}^{{\rm{odd}}}(x,y) \cdot \sin (o)} } }
\end{array}} \right.
\end{equation}
where ${\bf{LG}}_{s,o}$ denotes an odd-LogGabor filter with scale $s$ and orientation $o$. The odd-LogGabor responses ${\bf{G}}_x^{{\rm{LG}}}(x,y)$ and ${\bf{G}}_y^{{\rm{LG}}}(x,y)$ is taken as the representation of gradients and in average squared gradient (ASG) calculation:
\begin{equation}
\left[ \begin{array}{l}
{{\bf{G}}_{{W_\sigma },s,x}}\\
{{\bf{G}}_{{W_\sigma },s,y}}
\end{array} \right] = \left[ \begin{array}{l}
\sum\limits_{{W_\sigma }} {{{({\bf{G}}_x^{{\rm{LG}}})}^2} - {{({\bf{G}}_y^{{\rm{LG}}})}^2}} \\
\sum\limits_{{W_\sigma }} {2{\bf{G}}_x^{{\rm{LG}}}{\bf{G}}_y^{{\rm{LG}}}}
\end{array} \right]
\end{equation}

\renewcommand{\arraystretch}{0.9}
\begin{table*}[!h]\footnotesize
  \centering
  \setlength{\tabcolsep}{1.5mm}
  \caption{Average NCMs of multi-modal image matching results comparison by six methods.}
  \label{tbl:result}
    \begin{tabular}{cccccccccc}
    \toprule
    \multirow{2}{*}{Method} & \multicolumn{6}{c}{Remote sensing} & & \multicolumn{2}{c}{Medical}\\ \cline{2-7} \cline{9-10}
                            & optical-optical & optical-infrared & optical-depth & optical-SAR & optical-map & others & & optical & others\\

    \toprule
    SIFT         & 21.6 & 2.8 & 0.9 & 1.1 & 0.1 & 7.9 & & 10.6 & 0\\
    PSO-SIFT     & 14.9 & 50.7 & 3.0 & 0 & 1.0 & 14.2 & & 8.9 & 0\\
    PIIFD        & 44.0 & 34.0 & 11.4 & 12.3 & 2.1 & 2.0 & & 6.5 & 0\\
    MS-HLMO      & 208.1 & \bf{270.3} &  75.4 & 112.7 &  91.1 & 108.2 & & 97.0 & 0\\
    Proposed     & \bf{227.2} & 269.3 &  \bf{89.0} & \bf{147.7} & \bf{101.1} & \bf{109.8} & & \bf{183.5} & \bf{11.0}\\
    RIFT         & 390.5 & 279.3 & 294.0 & 263.8 & 255.0 & 160.3 & & 157.0 & 3.1\\
    Proposed$^+$ & \bf{523.2} & \bf{447.0} & \bf{340.9} & \bf{349.2} & \bf{382.4} & \bf{275.8} & & \bf{317.0} & \bf{93.0}\\
    \bottomrule
    \end{tabular}
\end{table*}

Another issue is that, according to visual saliency, the pixel closer to the center pixel has more importance. So an improvement is made to the PMOM that in a local area, features from a larger scale should have fewer effects. Each scale is given a weight that is inversely proportional to the scale, then the local orientation is calculated as:
\begin{equation}
{\bf{G}}_{{\rm{PMOM}}}^{{\rm{LG}}} = \frac{1}{2}\angle (\sum\limits_\sigma  {\frac{1}{\sigma }{{\bf{G}}_{{W_\sigma },s,x}}} ,\sum\limits_\sigma  {\frac{1}{\sigma }{{\bf{G}}_{{W_\sigma },s,y}}} )
\end{equation}
\begin{equation}
\angle (X,Y) = \left\{ \begin{array}{l}
\arctan (\frac{Y}{X}),X \ge 0\\
\arctan (\frac{Y}{X}) + \pi ,X < 0,Y \ge 0\\
\arctan (\frac{Y}{X}) - \pi ,X < 0,Y < 0
\end{array} \right.
\end{equation}
This orientation feature map ${\bf{G}}_{{\rm{PMOM}}}^{{\rm{LG}}}$ based on odd-LogGabor and weighted-PMOM (WPMOM) has much stronger invariance and stability in multi-modal images. The schematic diagram of feature extraction process is shown in Fig\ref{fig:wpmom}.

\subsection{Feature points description and matching}
\label{ssec:subhead}

After the most crucial invariant feature map is obtained, the next step is to extract a descriptor for each keypoints with a selected structure. Theoretically, any kind of valid descriptor structure is feasible. In the proposed framework, the generalized gradient location and orientation histogram-like (GGLOH) feature descriptor \cite{gao2022multi} is adopted for feature description, which has proved a better performance and is more operable. The WPMOM value at each keypoint is taken as the reference orientation, and the WPMOM values within the local area are counted with GGLOH to obtain descriptor vectors of the keypoints which are then for matching.

To deal with scale differences, the feature extraction is extracted in the scale-space by using Gaussian pyramid. The multi-scale strategy proposed in \cite{gao2022multi} has shown effective results, which is adapted to the registration process. The commonly used nearest neighbor (NN) matching and fast sample consensus (FSC) outlier removal are utilized for the proposed algorithm and also for each method in the experiments to get a fair comparison.

\section{EXPERIMENTS AND ANALYSIS}
\label{sec:experiment}

To comprehensively test the proposed method, the data sets provided in \cite{yao2022multi}, and \cite{gao2022multi} are adopted, and we also carefully prepared multi-source images from a broader field, which includes remote sensing, medical, and other natural or artificial images. The sensor types include optical, infrared, depth, synthetic aperture radar (SAR), digital map, etc. SIFT \cite{lowe2004distinctive}, PIIFD \cite{chen2010partial}, PSO-SIFT \cite{ma2016remote}, RIFT \cite{li2019rift}, and MS-HLMO \cite{gao2022multi} are selected for comparison. The number of correct matches (NCM) is employed as the evaluation metric. The experiments are all implemented using MATLAB2021b on a Windows with Intel Core I7-8700 CPU.

\begin{figure}[!h]
    \centering
        \subfigure[HSI-MSI]{\includegraphics[height=0.55in]{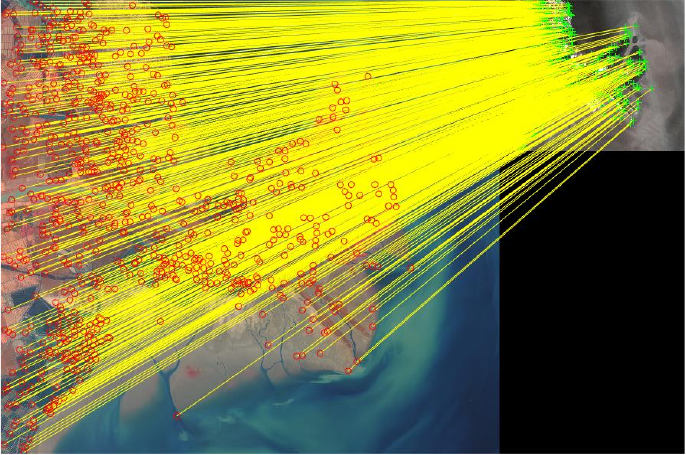}
        \label{fig:matching:a}}
        \subfigure[optical-infrared]{\includegraphics[height=0.55in]{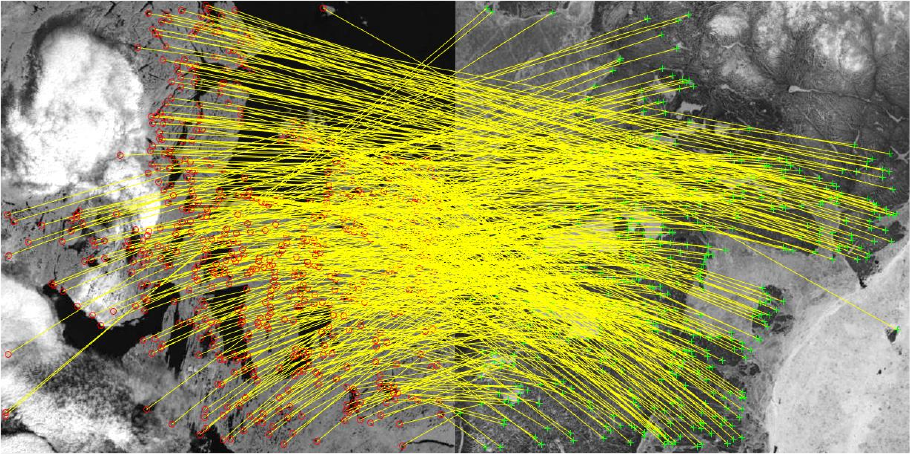}
        \label{fig:matching:b}}
        \subfigure[optical-depth]{\includegraphics[height=0.55in]{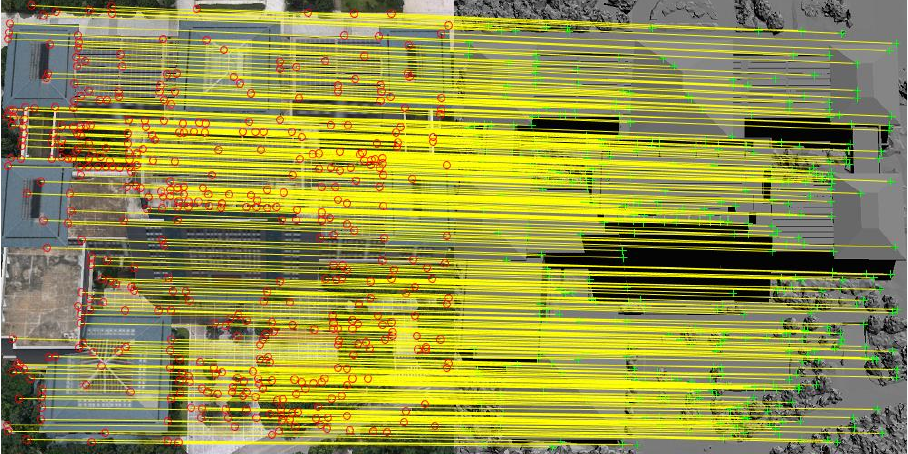}
        \label{fig:matching:c}}\\
        \subfigure[optical-SAR]{\includegraphics[height=0.50in]{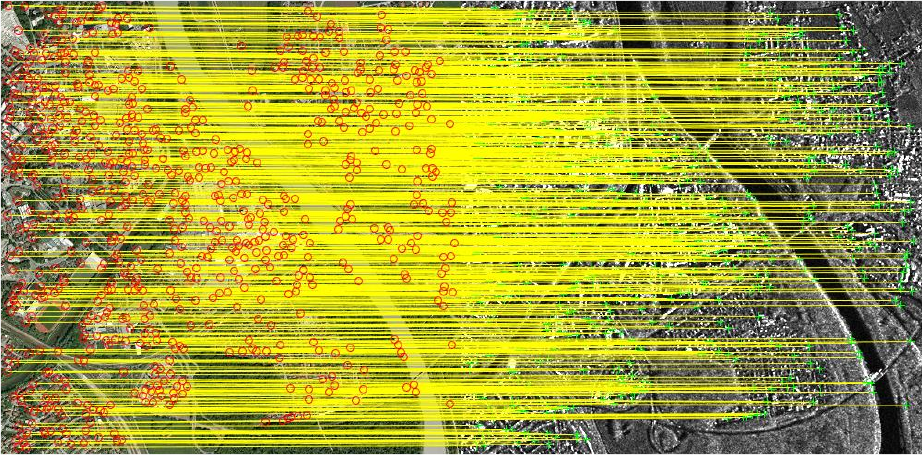}
        \label{fig:matching:d}}
        \subfigure[optical-map]{\includegraphics[height=0.50in]{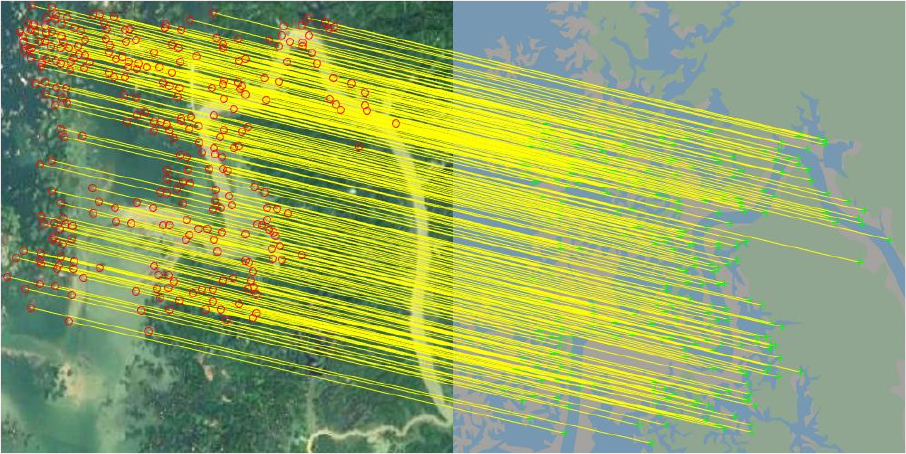}
        \label{fig:matching:e}}
        \subfigure[SAR-SAR]{\includegraphics[height=0.50in]{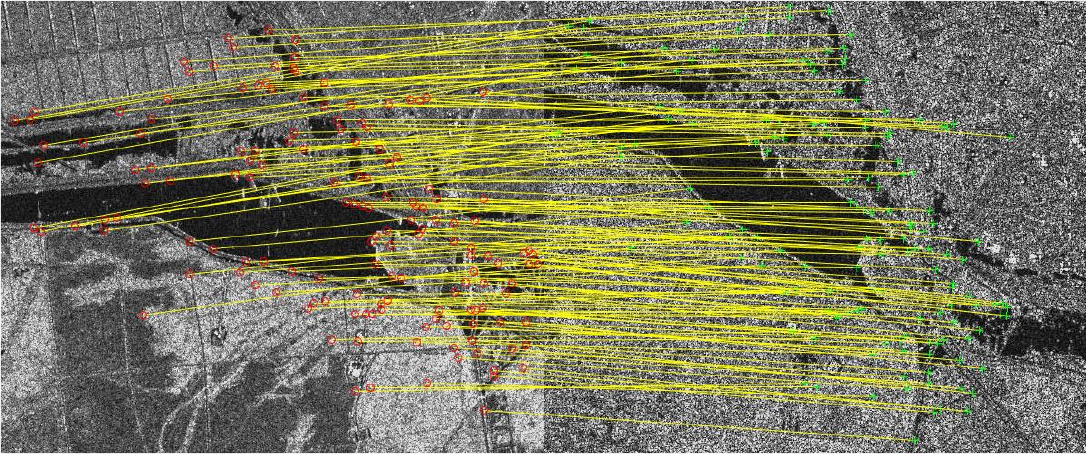}
        \label{fig:matching:f}}\\
        \subfigure[tissue fluorescence]{\includegraphics[height=0.58in]{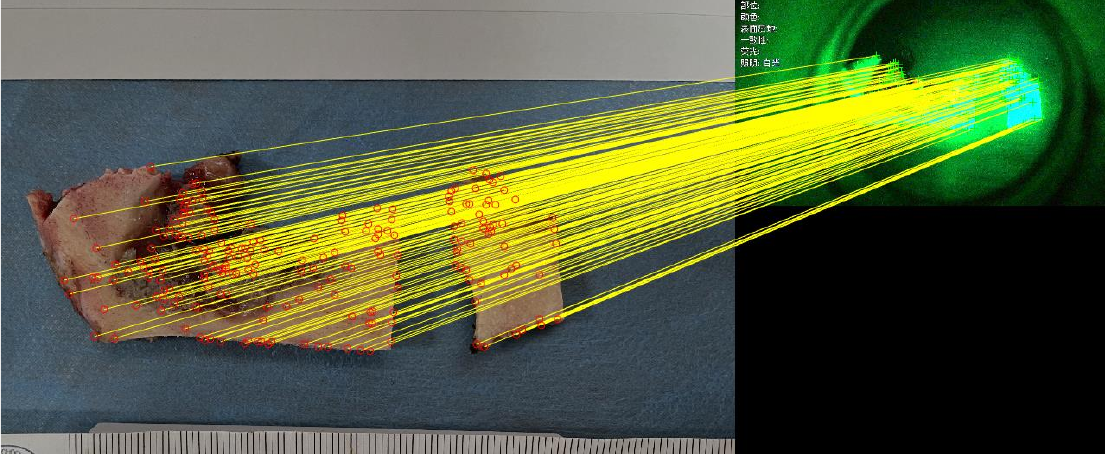}
        \label{fig:matching:g}}
        \subfigure[staining]{\includegraphics[height=0.58in]{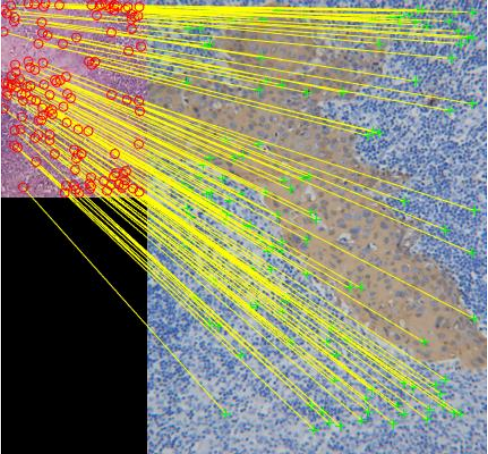}
        \label{fig:matching:h}}
        \subfigure[retina fundus]{\includegraphics[height=0.58in]{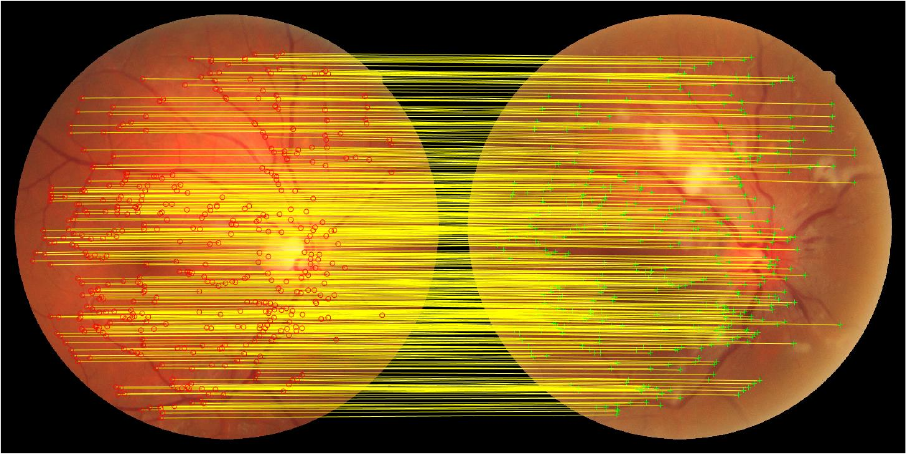}
        \label{fig:matching:i}}
    \caption{Examples of multi-modal images matching from remote sensing and medical fields.}
    \label{fig:matching}
\end{figure}

The average NCM of each group of multi-modal image pairs are listed in Table \ref{tbl:result}. Note that, due to RIFT's treatment to rotation property being more of an enumeration, adjustments are made accordingly to the proposed method for a fairer comparison, applying the same strategy, marked as Proposed$^+$. It shows that the proposed method obtains the most NCMs in almost all groups. The performance of MS-HLMO is the best among compared algorithms, whose average NCM exceeds the proposed method in one group. Through a detailed analysis, it is found that MS-HLMO has very high NCMs in some image pairs, which pulls up the average, while it even fails in some others. The proposed method's effect is satisfactory on all images. Therefore, it shows that the proposed method has better robustness and generalization, which is more unaffected by the data modal. Fig.\ref{fig:matching} shows some examples of the keypoint matching results.

Through inspecting the transformed and aligned images and comparison with manual matching results, the spatial losses of matched image pairs are within 1$\sim$2 pixels. The proposed algorithm is proved to be effective.

\section{CONCLUSION}
\label{sec:conclusion}

In this paper, an effective feature-based image matching algorithm is proposed for multi-modal images. The invariant features are analyzed and summarized, providing a robust feature based on local gradient orientation and LogGabor filter. Through experiments on a comprehensive set of multi-source data, the algorithm can well achieve the matching tasks, where the image matching losses can reach within 1$\sim$2 pixels. It is concluded that the algorithm has good robustness, stability, and generalization.

\bibliographystyle{IEEEbib}
\bibliography{refs}

\end{document}